\documentclass[12pt,preprint]{aastex}
\begin{document}

\newcommand{\kms}{\mbox{km~s$^{-1}$}}
\newcommand{\s}{\mbox{$''$}}
\newcommand{\mloss}{\mbox{$\dot{M}$}}
\newcommand{\my}{\mbox{$M_{\odot}$~yr$^{-1}$}}
\newcommand{\ls}{\mbox{$L_{\odot}$}}
\newcommand{\ms}{\mbox{$M_{\odot}$}}
\newcommand\mdot{$\dot{M}  $}
\title{BINARITY IN COOL ASYMPTOTIC GIANT BRANCH STARS: A GALEX SEARCH FOR ULTRAVIOLET EXCESSES}

\author{R. Sahai\altaffilmark{1}, K. Findeisen\altaffilmark{2}, A. Gil de Paz\altaffilmark{3}, C.
S{\'a}nchez Contreras\altaffilmark{4}
}

\altaffiltext{1}{Jet Propulsion Laboratory, MS\,183-900, Caltech,
    Pasadena, CA 91109}
\altaffiltext{2}{Cornell University, Ithaca, NY 14853}
\altaffiltext{3}{Dpto. de Astrof\'{\i}sica, Universidad Complutense de Madrid, Madrid
28040, Spain}
\altaffiltext{4}{Dpto. de Astrof\'{\i}sica Molecular e
Infraroja, Instituto de Estructura de la Materia-CSIC, Serrano 121, 28006 Madrid, Spain}

\authoremail{sahai@jpl.nasa.gov}

\begin{abstract}
The search for binarity in AGB stars is of critical importance for our
understanding
of how planetary nebulae acquire the dazzling variety of aspherical shapes which
characterises this class. However, detecting binary companions in such stars has
been severely hampered due to their extreme luminosities and pulsations. We have
carried out a small imaging survey of AGB stars in ultraviolet light (using GALEX)
where these cool objects are very faint, in order to search for hotter companions.
We report the discovery of significant far-ultraviolet excesses towards
nine of these stars. The far-ultraviolet excess most likely results either directly from the
presence of a hot binary companion, or indirectly from a hot accretion disk around the
companion.

\end{abstract}


\section{Introduction}

There are many observational indications which lead us to believe that binarity,
believed to be very common amongst pre-main-sequence (e.g., Bodenheimer et
al. 2000) and main-sequence stars (Duquennoy \& Mayor 1991), strongly influences the history
and geometry of mass loss during the
late stages of stellar evolution. The evolutionary transition from the AGB to the
post-AGB phase is accompanied by significant changes in the morphology of these
objects -- the roughly round circumstellar mass-loss envelopes (CSEs) of AGB
stars evolve into post-AGB nebulae with a dazzling variety of shapes and
intriguing symmetries (e.g$.$ Schwarz et al$.$ 1992, Sahai \& Trauger 1998).
Critical reviews (Soker 1998) of the properties of bipolar PNe (e.g$.$ Corradi \&
Schwarz 1995) lead to the conclusion that binary models can explain all these
properties, whereas single-star models (e.g$.$ Garc\'{\i}a-Segura 1997) have many
difficulties.

However, in spite of dedicated efforts by many researchers to search for binarity in evolved
stars, direct observational evidence for binarity has been hard to come by. AGB stars are
very luminous ($\sim\,few\times10^3$-10$^4$\,\ls) and surrounded by dusty envelopes, making
it very difficult to directly detect nearby stellar companions which are generally likely
to be significantly less luminous main-sequence stars or white dwarfs. Indirect techniques
such as radial-velocity measurements (e.g., van Winckel et al$.$ 1999, Sorensen \& Pollaco
2003, de Marco et al$.$ 2004) or photometric variability measurements (Bond 2000) have been
used for the central stars of PNs and post-AGB objects, with some success. But these
techniques cannot be easily applied to AGB stars, because the latter show strong 
variability intrinsic to their pulsating atmospheres, which potentially masks the
corresponding variability due to a companion. Extensive observations of the central stars
of planetary nebulae have resulted in detections a sum total of $\lesssim$20 binaries (Bond
2000, Ciardullo et al. 1999), implying a 10-15\% fraction of detectable close binaries
among randomly selected PNe. Bond (2000) concludes that it is likely that the known
short-period binaries in PNe are only the tip of an iceberg of a substantial population of
longer-period binaries.

Deep ultraviolet observations hold the promise of being able to discover
substantial numbers of binary companions in AGB stars, since most mass-losing AGB
stars are relatively cool objects (spectral types $\sim$M6 or later). The companions are 
likely to be main-sequence stars because of the steep dependence of evolutionary rates on stellar
mass 
(e.g. Soker \& Rappaport 2000). Thus for a secondary-to-primary mass ratio, $q=M_2/M_1$, around
unity, any
stellar companion has a good probability of being hotter than the primary. However, it is
difficult to estimate with confidence the number of such systems where the secondary is on the
main-sequence and hotter than the primary, as a fraction of the total number of primordial
binaries, since the mass-ratio probability distribution, $f(q)$, where $q=M_2/M_1$, is not
well-known. A promising approach is to carry out population synthesis studies (which are still
in their infancy) such as those of Soker \& Rappaport (2000), who adopt, for their modelling,
$f(q)\propto q^{1/4}$; note that this function is not strongly peaked towards $q=1$.

Since observed and model spectra of cool AGB stars show that their fluxes die rapidly at
wavelengths shortwards of about 2800\AA, significantly favorable secondary-to-primary flux
contrast ratios ($>$10) for companion detection may be reached in the GALEX FUV 
(1344-1786\AA) and NUV (1771-2831\AA) bands, for companions of spectral type hotter than
about G0 ($T_{eff}$=6000\,K). In this paper,
we report on a subsample of objects from our Cycle 1 pilot program, which were detected in
both the FUV and NUV bands, and the implications of these detections for binarity. A
comprehensive study covering the
full results of our survey will be presented in a forthcoming paper.

\section{Observations \& Results}
We selected a sample of 25 AGB late-M (i.e., M5 or later) stars (which passed the GALEX
mission ``bright-star" and ``high-background" tests) largely based on their inclusion in
the Hipparcos astrometric catalog, with a 
``multiplicity" flag in the header field H59 of the main catalog, 
indicating that a single-star astrometric
solution was not adequate\footnote{these were the so-called problem stars for which a double star
solution with the classical parameters of separation and position angle could not be found: and
were flagged as G, O, V, or X in field H59 of the main catalog}. Thus for 20/25 objects,
the selection criteria of our ``pilot" program were intentionally biased towards optimising
the {\it a priori} probability of finding companions, in order to test the validity of our
technique. Three objects did not have a ``multiplicity"
flag; and two are not in the Hipparcos catalog, but were selected from published lists of AGB
stars with molecular envelopes detected in CO emission. Although most of our objects had positive
entries for the annual parallax, the errors were usually large, and only for four objects were
the parallax measurements significant (i.e., greater than 3$\sigma$). The requirement that our
objects 
be M5 (or later) or cool N-type carbon stars was included in order to minimise the ratio of 
their UV fluxes to that of hotter companions (if present).  

From our original list of 25 objects, 21 objects have been observed -- 20 as part of our
GI program (GI1-23; PI: R$.$Sahai), and 1 as part of other programs -- in both the NUV
(1771-2831\AA) and FUV (1344-1786\AA) bands. Nine of these were detected in both the FUV and
NUV bands with high S/N ($\gtrsim\,8\sigma$) (Table\,\ref{fluxes}). Amongst these, the NUV
image of AF\,Peg has an elliptical shape, and an intensity cut along the major axis
(PA$\sim75\arcdeg$) shows two peaks separated by about 6$''$; the stronger one corresponds
to AF\,Peg's location; in the FUV image, AF\,Peg is weaker. Because the separation is
comparable to the PSF, the two sources cannot be deconvolved reliably; hence the measured
fluxes of AF\,Peg are very uncertain. 
We have used the pipeline-generated catalogs included with the
imaging datasets for extracting the photometry for the remaining 8 sources. Stars not
detected in the FUV, will be discussed together with their NUV properties, in a forthcoming
paper.
  
The 8 FUV sources include 4 oxygen-rich stars (RW\,Boo, AA\,Cam, V\,Eri \& R\,UMa) and 4
carbon-rich
stars (T\,Dra, TW\,Hor, V\,Hya, \& VY\,UMa). 
For those objects where more than one exposure was available, taken at different epochs, we
list both the individual and average fluxes.
The typical uncertainty in the measured fluxes is dominated by systematic
uncertainties in the GALEX pipeline photometric calibration of about 10-15\% (Morrissey et al.
2005).

We now consider whether our FUV and/or NUV detections could result from the presence of a
small filter red leak, which could produce 
spurious detection of a UV signal for the extremely red stars observed in this study. Since the
GALEX detectors are photon counting MAMA detectors, there is supposedly no ``red leak", since
only UV photons can trigger the photoelectrons (Rich 2005); the photocathode on the FUV
detector is non-responsive above $\sim$1800\,\AA, and the NUV response is suppressed below
measurable levels by multilayer coatings on the optics. According to the GALEX helpdesk, there is
no measurable red leak in either GALEX band. 

Even though no red leak response has been measured for the GALEX filters, we have ensured that
even if such a response is present at a low level, our modelling is not affected by the latter
because the upper limits which we can set on the red leak from our data are quite low. We have
done this by comparing, for our survey objects, the ratios of the FUV and NUV fluxes to the
V-band fluxes and assuming that the lowest of these ratios are due to a red leak. From this
analysis, we find  values of $2.5\times10^{-7}$ and $4\times10^{-6}$ for the maximum possible
``red leak" flux in the GALEX FUV and NUV bands, as a fraction of the V-band flux -- these
ratios are too low to affect our models, or the detection statistics we report in this paper.


\section{Ultraviolet Excesses}

\begin{figure}[htb]
\vskip -0.2in
\resizebox{1.0\textwidth}{!}{\includegraphics{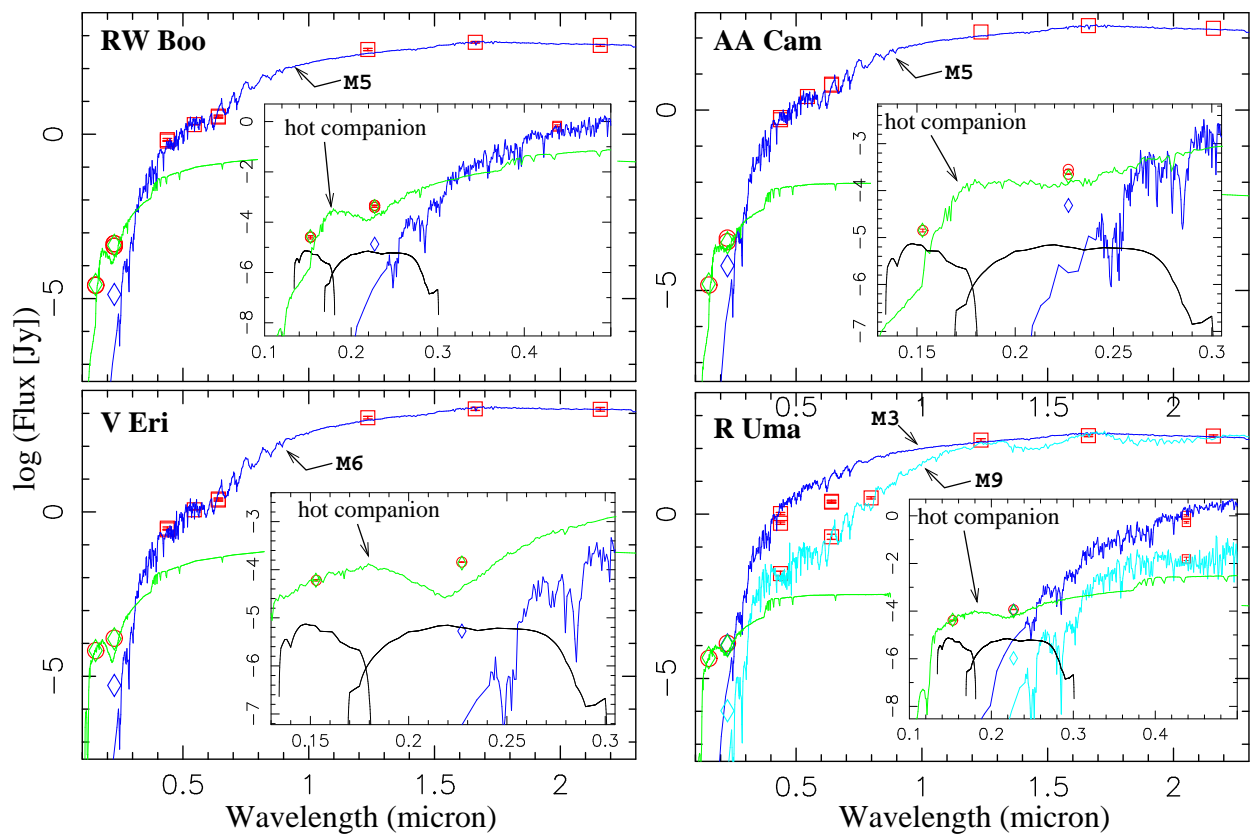}}
\vskip -0.1in
\caption{GALEX (NUV \& FUV) and ground-based (optical \& near-IR) fluxes ({\it red
symbols}) of AA Cam, V Eri \& R\,UMa, with model spectra ({\it blue}: cool AGB
star, {\it green}: hot companion). The expected NUV flux ({\it blue/cyan diamonds}) from the
cool AGB star (the expected FUV flux due to AGB star lies below the minimum of the flux
range), and from the hot companion + cool AGB star ({\it green diamonds}) are also shown. 
The inset shows an expanded view of the UV-Blue region. 
The model NUV/FUV fluxes ({\it diamond symbols}) are obtained by convolving the model
spectrum with the GALEX NUV and FUV bandpasses ({\it black curves, inset}).
}
\label{sed-fit}
\end{figure}

We now investigate the origin of the ultraviolet fluxes in the objects we have detected in the
FUV band. We have fitted the spectral-energy-distributions (SEDs) from 0.1 to 2\micron
(Fig.\ref{sed-fit}) of the four oxygen-rich stars with reliable FUV fluxes (i.e. RW\,Boo,
AA\,Cam, V\,Eri and R\,UMa), using stellar atmosphere models of AGB stars (Fluks et al. 1994),
corresponding to the spectral type of each star as given in the General Catalogue of Variable
Stars (GCVS: Samus et al. 2004\footnote{available online through the Vizier astronomical
database}). A visual extinction, $A_V$, to account for the extinction by circumstellar dust due
to the dusty mass-loss envelope of the primary, is also determined from our fits. Archival
photometry at wavelengths redwards of the GALEX NUV band was taken from the Hubble Guide Star
Catalog (GSC 2.2), the US Naval Observatory USNO-B1.0 Catalog, and the Two Micron All-Sky
Survey.


We find that the observed FUV (NUV)
fluxes are a factor $>10^{6}$ ($>$5) larger than expected for the photospheric 
emission of the primary, accounting for the
finite filter bandwidth, the filter response\footnote{taken from
http://galexgi.gsfc.nasa.gov/docs/galex/tools/Resolution\_Response/index.html}, and the steeply
sloping spectrum in the UV.
Hence, even though the photometric variability of the
primary stars makes our fit somewhat uncertain, it certainly cannot account for
the FUV excesses, because they are very large, and there is no
systematic relationship between the light-cycle phases of the various photometric
data-points used to fit the primary model which could conspire to produce such an
excess in each of our three sources. 

The detailed SED fitting described could not be carried out for the 4 carbon-rich stars
in Table\,\ref{fluxes} because, for these objects, model atmospheres for
$\lambda\lesssim2300$\AA~are not available (D. Luttermoser, priv.comm.). We therefore used
black-body spectra using $T_{eff}$ values from Bergeat et al. (2001), and scaled these to
fit the NIR and optical photometry of each object. We find that the model FUV flux of the
primary AGB star, even for the object with the smallest FUV excess (VY\,UMa), is lower than
the observed value by a factor $\sim$30. V\,Hya stands out amongst the 9 FUV-detected stars
as having the highest FUV-to-NUV flux ratio ($\gtrsim$\,1), but the coolest photosphere
($T_{eff}=2160$\,K) for the primary. We discuss our detection of the FUV/NUV excess in this
object in more detail later (\S\,\ref{discuss}).

We note that Wood \& Karovska (2004) conclude,
based on IUE spectroscopic data of several Mira AGB stars at multiple phases, that this class of
objects do not produce any detectable emission below 2000\AA.
We have examined the IUE database for the sources in our survey, and find that only 3 objects in
our survey sample were observed -- R LMi, V Hya and TW Hor. R LMi and TW Hor were observed with
both the long-wavelength (LWR and LWP) and short-wavelength (SWP) instruments (Boggess et al.
1978a,b), whereas V Hya was observed only with the long-wavelength instrument. For both R Lmi
and V Hya, the signal-to-noise ratio over the observed bandpasses (1910--3300\,\AA~for LWR/LWP
and 1150--1975\,\AA~for the SWP) is close to zero. For TW Hor, the situation is similar, except 
at wavelengths longer than $\sim$2500\,\AA, where significant flux is detected; the steady rise
in the spectrum towards the red end of the bandpass indicates that this flux is most likely due
to the primary star. We have convolved the IUE spectra with the GALEX FUV
and NUV filter bandpasses in order to compute the GALEX-equivalent fluxes (or upper limits) for
these sources. We find, for R Lmi, 3$\sigma$ upper limits of 0.032 mJy (FUV) and 0.16 mJy (NUV).
For the other two sources, the derived fluxes (3$\sigma$ errors) are as follows: TW Hor --
$0.096\pm0.051$ mJy (FUV) and $0.41\pm0.05$ mJy (NUV); V Hya -- $0.12\pm0.11$ mJy (NUV).


We now examine two plausible explanations for the FUV excesses, both of which involve the
presence of a companion star.

\subsection{A Hot Companion}

The NUV and FUV excesses may result from the presence of a
companion star which is significantly hotter than the primary. 
We have therefore made least-squares fits to the FUV and NUV excesses of each
object by including the contribution of a companion star (Table\,\ref{models}).
We have used models by Kurucz (Castelli \& Kurucz 2002) for the companion spectra.

For RW\,Boo, AA\,Cam and V\,Eri, the same value of $A_V$ was applied to the model
spectrum of the companion, as derived from fitting the primary. In the case of 
R\,UMa, which is listed in GCVS with a spectral type M3-M9, the best fits to the 
optical and near-infrared fluxes were obtained with  $A_V=2.6\,(0.0)$ for an M3 (M9) primary
spectrum. Therefore, in our least-squares fitting of the companion, we tried three values of  
$A_V$ (0,1.3 \& 2.6) to scale the companion black-body spectrum.

Our modelling (Table\,\ref{models}) provides the fractional luminosity 
(relative to the primary) of the companion. For RW\,Boo and V\,Eri, the
Hipparcos parallaxes, $3.09\pm1.1$\,mas and $4.56\pm1.08$\,mas, give distances of
320 and 220\,pc, respectively, implying companion
luminosities of, $L_c$= 18 and 6\,\ls~for these two sources. For AA\,Cam and R\,UMa, the
parallax measurements are not
significant, and the luminosities given in Table\,\ref{models} are for a nominal
distance of 0.5\,kpc. 

The value of $L_c$ for RW\,Boo is consistent with that expected for a mid-A 
main-sequence star, but it is too low in the case of AA\,Cam, R\,UMa, and V\,Eri, since the 
luminosities of main-sequence stars using our most favorable (i.e. lowest) model values for each
of these, 
$T_{eff}=7800, 8900, 9250$\,K (i.e., spectral type $\sim$A6 to $\sim$A1-A2),  
lie in the range $\sim$(10--35)\,\ls~(Cox 2000, Table 15.7). 
An appeal to distance ambiguities for these 3 stars does not help to resolve the problem of
the derived $L_c$ values being too low for 
main-sequence stars. We have tried increasing the source distance in order to bring up $L_c$
to its main-sequence values. For AA\,Cam and R\,UMa, this
exercise results in $L_p$=$2.4\times10^4$\ls~and $9.3\times10^4$\ls, using the most favorable
models values in Table 2 - i.e., 
the lowest values of $T_{eff}$ (7800\,K and 8900\,K) and the highest values of
$L_c$ (1.79 and 1.01\,\ls). The value of $L_p$ for R\,UMa (AA\,Cam) is 
certainly (probably) too high for an AGB star. For V\,Eri, a factor 2.4 increase in
the distance is needed to scale up $L_c$ to a main-sequence luminosity, but such a large
increase is significantly beyond that allowed by its parallax data (factor 1.3). We rule out
the possibility that the companions are low-luminosity white dwarfs (WDs) on cooling tracks,
because stellar evolutionary models show that by the time WDs have cooled to $10^4$\,K,
their luminosities are orders of magnitude below 1\ls.



In our models we adopted the extinction curve as tabulated by Whittet (1992). Our quoted
modeling uncertainties do not take into account uncertainties in the extinction curve at NUV
and FUV wavelengths. We repeated the modeling using extinction curves for the LMC Supershell
and the SMC Bar (Gordon et al. 2003), which along with the Galactic extinction roughly cover
the range of curves found in circumstellar dust and are well-studied. 
The results are shown in Table~\ref{dusteffect}. Although the best-fit temperatures all shifted
up when the LMC and SMC extinction curves were used, the cooler companion models (i.e., for
RW\,Boo and AA\,Cam) proved reasonably insensitive to the choice of extinction curve, but
dramatic differences were found for R\,UMa and V\,Eri. In general, we could not obtain a good
fit even with the highest-temperature models available (39000\,K) while using the LMC or SMC
curves for these two sources.

%
%

\subsection{Accretion onto a Companion Star}
\label{accrete}
Five out of nine objects in Table\,\ref{fluxes} were observed on more than one epoch in one or
both of
the GALEX bands -- in each instance, significant photometric variability was observed
(Table \,\ref{fluxes}). We have checked that this variability is not due to systematic
calibration uncertainties because the average and median fractional differences of the
fluxes for the brightest 40 field objects in the images from the different epochs are
negligible.

A plausible interpretation for the photometric variability is related to the
presence of a nearby companion. This interpretation is motivated by ultraviolet observations
of Mira, a symbiotic star in which Mira\,A is the AGB
primary and Mira\,B is a compact companion (at a separation of 0\farcs6) which is accreting
matter from Mira\,A's wind. IUE spectra of Mira in
the wavelength region covered by the GALEX bands show the presence of strong emission
lines ascribed to Mira B (Reimers \& Cassatella 1985). The strongest of these (due to O,N,
\& C, seen by IUE during 1979-95) were found to fade by a factor $>$20 by  
1999-2001, and then start increasing back to their original levels by 2004 (Wood \& Karovska
2006). Assuming the revised Hipparcos-based distance of 107\,pc to Mira, the combined maximum
fluxes of
such emission lines, if
present in our sources, would correspond to an artificial continuum in GALEX's broad-band
FUV filter of about 0.1 mJy at a typical source distance of 350\,pc, thus comparable to the
measured values of the FUV fluxes in our sources. During the IUE era
(1979-80 \& 1990-95), the FUV lines and continuum varied by a factor $\sim$2 in Mira.
Karovska, Wood and co-authors (Wood, Karovska \& Hack 2001, Wood, Karovska \& Raymond 2002)
conclude that the UV variability most likely results from variations in the accretion rate
onto Mira\,B.
Although Mira's variability has been observed on a much longer time-scale than the ones
sampled in our GALEX data, accretion of matter from the primary AGB wind onto a companion
provides a plausible explanation for the presence of FUV emission and its variability in
our sources.

\section{Discussion}
\label{discuss}
Our small survey of 21 AGB stars for UV excesses has resulted in a substantial
number of NUV and/or FUV detections. Nine of these were detected in the FUV band and are
the subject of this paper. A detectable FUV flux at even a few $\mu$Jy level is several
orders of magnitude too high to be explained by photospheric emission from the relatively
cool primary stars in our sample, and hence is an ``excess" which requires an alternative
explanation -- most likely the presence of a binary companion. The excesses arise either as a
result of photospheric emission from a hotter companion, and/or from an accretion disk around
the companion. Spectroscopic monitoring in the FUV of these sources is needed in order to
distinguish between these two mechanisms. 

We detected NUV fluxes in
19/21 of our objects with high S/N, many of which are also likely to be ``excesses", but
for which such an inference is more uncertain because of the significantly larger
contribution of the primary in the NUV compared to the FUV. A discussion of the detection
statistics of, and the biases in, our full sample, is deferred to a forthcoming paper.  

Although our discovery of a UV-excess attributable to a different star than the primary AGB
star does not directly imply that the former is a gravitationally-bound companion, it is
the most likely explanation. This is because the UV sky is rather ``empty" (i.e., much more
scarcely populated than at optical wavelengths), hence the probability, $p_{false}$ that the
FUV-emitting object is simply positionally coincident on the sky with the primary (i.e.,
lying within a radius of 2$''$ from the primary), is very small. Using the object number
count (per deg$^2$ mag) versus magnitude plot for the GALEX FUV band (Bianchi et al. 2007),
we find that for AA\,Cam (the faintest of our modelled sources), $p_{false}$ for an object 
of FUV magnitude lying within a 0.5 mag bin centered around the FUV mag of AA\,Cam, is
$8\times10^{-5}$. The Galactic latitudes of our objects are similar to those of the fields used
by Bianchi et al., hence it is appropriate to use their point-source densities.

V\,Hya, which has the largest FUV flux, as well as the highest FUV-to-NUV flux ratio amongst
all our targets, is well known for its collimated, high-velocity, outflows, an extended dusty
torus, and an inner hot disk. The outflows were first seen via infrared
absorption lines in the CO
4.6\,\micron~vibration-rotation band (Sahai \& Wannier 1988); recent interferometric mapping of
the millimeter-wave CO line emission shows the collimated structure of the fast outflow (e.g.,
Hirano et al. 2004). More recently, observations with
STIS/HST reveal the presence of a high-velocity blob moving away from the central source at
(projected) speeds upto 220\,\kms, and a hot, slowly expanding (10--15\,\kms) central disk-like
structure (Sahai et al. 2003). Although the expansive kinematics of the latter implies that it
is not an accretion disk, it may result from a recent phase of equatorially enhanced mass-loss,
which may be enhancing the accretion process. 
V\,Hya is thus the best example to date of an evolved star with an active, collimated outflow,
dense equatorially-flattened structures possibly related to a central accretion disk, and an
inferred binary companion from our UV excess measurements.

\acknowledgements
We would like to thank an anonymous referee for his/her thoughtful review of our paper. We
acknowledge discussions with Patrick Morrissey related to the possibility of red leaks in the
GALEX FUV and NUV bands.  RS's contribution to the research described in this publication was
carried out at the Jet Propulsion Laboratory, California Institute of Technology, under a
contract with NASA. RS thanks NASA for financial support via a GALEX award and an LTSA
award. KF was partially funded by a SURF scholarship and through the Cornell Presidential
Research Scholars (CPRS) program.AGdP is partially financed by the Spanish {\it Ram\'{o}n y
Cajal} program and the {\it Programa Nacional de Astronom\'{\i}a y Astrof\'{\i}sica} under grant
AYA 2006-02358. CSC is partially funded for this work by the Spanish MCyT under project AYA
2006-14876 and the Spanish MEC under project PIE 200750I028. 


%
\begin{table}
\scriptsize
\caption{AGB Stars with UV Excesses\label{fluxes}}
\begin{tabular}[]{|c|c||l|c|c|c}
\tableline
Target  & Band & Epoch\tablenotemark{a} & Exp.Time (s) & Flux (mJy) \\
\tableline
RW Boo\tablenotemark{b}& FUV & 4220.75 & 1726  & 0.026 \\ 
        & NUV & 3861.15 & 3396  & 0.47 \\
        &     & 4220.75 & 1726  & 0.39 \\
        &     & Average & 5122 & 0.44  \\
\tableline
AA\,Cam & FUV & 3377.48 & 1693 & 0.014 \\
        & NUV & 3377.48 & 1532 & 0.22  \\
        &     & 3425.29 & 1693 & 0.28  \\
        &     & Average & 3225 & 0.25  \\
\tableline
T Dra   & FUV & 3984.2 & 1580 & 0.0055\\ 
        & NUV & 3541.9 & 1250 & 0.017 \\
        &     & 3587.5 & 1309 & 0.018 \\ 
        &     & 3984.2 & 1580 & 0.032 \\
        &     & Average & 4151 &  0.021 \\
\tableline
V\,Eri  & FUV & 3678.20 & 1620 & 0.060 \\
        & NUV & 3678.20 & 1704 & 0.14  \\
\tableline
TW\,Hor\tablenotemark{b} & FUV & 3349.3  & 4381 & 0.026 \\
        &     & 3351.2  & 1345 & 0.032 \\
        &     & 3671.8  & 109  & 0.034 \\
        &     & 3706.6  & 255  & 0.023 \\
        &     & 3706.7  & 586  & 0.033 \\
        &     & 4018.8  & 747  & 0.034 \\
        &     & Average & 7423 & 0.027 \\
        & NUV & 3349.3  & 4381 & 0.53 \\ 
        &     & 3351.2  & 1345 & 0.55 \\
        &     & 3359.25 & 2214 & 0.34 \\
        &     & 3671.8  & 109  & 0.78 \\
        &     & 3706.6  & 339  & 0.26 \\
        &     & 3706.7  & 645  & 0.27 \\
        &     & 4018.8  & 747  & 0.99 \\
        &     & Average & 9780 & 0.49 \\
\tableline
V\,Hya  & FUV & 3421.86 & 1705 & 0.12 \\
        &     & 3778.54 & 1003 & 0.15 \\
        &     & Average & 2708 & 0.13 \\
        & NUV & 3421.86 & 1705 & 0.11 \\
        &     & 3778.54 & 1003 & 0.13 \\
        &     & Average & 2708 & 0.12 \\
\tableline
AF\,Peg\tablenotemark{c} & FUV & 3280.70 & 1551 & 0.011 \\
        & NUV & 3280.70 & 1551 & 0.090 \\
\tableline
R\,UMa  & FUV & 3742.14 & 1703 & 0.041 \\
        & NUV & 3742.14 & 1703 & 0.12  \\
\tableline
VY\,UMa & FUV & 3742.50 & 1704 & 0.0061 \\
        & NUV & 3742.50 & 1704 & 0.23 \\
\tableline
\end{tabular}
\tablenotetext{a} {JD-2450000}
\tablenotetext{b} {Observed at multiple epochs; data from epochs separated by $\lesssim$0.1
day are averaged together in the individual-epoch rows}
\tablenotetext{c} {Observed at 2 epochs in the NUV band only; quoted
fluxes are for AF\,Peg using a 2-Gaussian fit to the image of the latter and the partially
blended nearby star, for Epoch 1}
\end{table}

\begin{footnotesize}
\begin{table}
\caption{Model Results\label{models}}
\begin{tabular}[]{|c|c|c|c|c|c|c}
\tableline
Target  & Primary & D &$A_V$ & $T_{eff}$\tablenotemark{a} & $L_c$\tablenotemark{a} &
$L_p/L_c$\\
        & Sp.Type & (kpc) &    &  (K)     &    (\ls) &          \\
\tableline
RW\,Boo & M5 & 0.32 & 2.3  &  8200 (-500,300)  & 18 (-5,7)      & 280 (-80,110) \\
AA\,Cam & M5 & 0.5  & 1.0  &  8200 (-400,400)  & 1.1 (-0.6,0.7) & 3200 (-1300,4500)\\
V\,Eri  & M6 & 0.22 & 2.9  &  10000(-700,4100) & 6.2 (-2.9,2.2) & 910 (-240,810)  \\
R UMa  & M3-M9&0.5  & 1.3  & 9200(-300,+1100) & 0.85(+0.2,-0.4) & 5300(-900,+4700)\\


\tableline
\end{tabular}
\tablenotetext{a}{The numbers in parenthesis represent 3$\sigma$ modelling uncertainties;
temperature and luminosity uncertainties are inversely correlated} 
\end{table}
\end{footnotesize}

 

\begin{table}
\caption{Best-fit temperatures (in K) under different extinction curves.\label{dusteffect}}
\begin{tabular}[]{|c|c|c|c|c|}
\tableline
Target	& $A_V$	& Galactic\tablenotemark{a} & LMC\tablenotemark{a} & SMC\tablenotemark{a} \\
\tableline
RW\,Boo	& 2.3	& 8200 (-500,300)	&  8700 (-500,400)	& No fit	\\
AA\,Cam	& 1.0	& 8200 (-400,400)	&  8500 (-400,500)	& 9000 (-400,1400) \\
V\,Eri	& 2.9	& 10000 (-750,4500)	& No fit		& No fit	\\
R\,UMa	& 1.3	& 9200(-300,+1100)	& 33000 (-22000)\tablenotemark{b}& No fit \\
\tableline
\end{tabular}
\tablenotetext{a}{The numbers in parenthesis represent 3$\sigma$ modelling uncertainties.} 
\tablenotetext{b}{The 3$\sigma$ upper limit for R\,UMa fell beyond the range of available
models.}
\end{table}




\clearpage

\end{document}